\documentclass[twocolumn,amsmath,amssymb]{revtex4}
\usepackage{graphicx}% Include figure files
\usepackage{dcolumn,setspace}% Align table columns on decimal point
\usepackage{bm}% bold math
\begin{document}
%\mark{{Gallant king...}{X Y zzz and A B zzzz}}
\title{Chaotic to ordered state transition of cathode-sheath instabilities in DC glow discharge plasmas}

\author{Md. Nurujjaman, and A.N. Sekar Iyengar}
\address{ 1/AF, Bidhannagar, Saha Institute of Nuclear Physics,India}
%\keywords{chathode-sheath, instabilities, chaos, period-subtraction, bifurcation, dc-discharge.}
%\pacs{}
\begin{abstract}
Transition from chaotic to ordered state has been observed during  the initial stage of  a discharge in a
cylindrical dc glow discharge plasma. Initially it shows a chaotic behavior but  increasing  the  discharge 
voltage changes the characteristics of the discharge glow and  shows a period substraction of order 7 period 
$\rightarrow$ 5 period $\rightarrow$3 period $\rightarrow$1 period i.e. the system goes to 
single mode through odd cycle subtraction. On further increasing  the discharge 
voltage, the system goes through period doubling, like 1 period $\rightarrow$ 2 period $\rightarrow$ 4 period.
On further increasing the voltage, the system goes to stable state without having any oscillations. 
\end{abstract}
\maketitle
\section{Introduction}
Nonlinear phenomena are abundant in nature and laboratory plasma. Plasma 
is a typical nonlinear dynamical system with a large number
of degrees of freedom, and a medium for testing a rich variety of
nonlinear phenomena such self oscillation, period doubling, bifurcation,
period subtracting, period adding, chaos, intermittency etc.
\cite{IEEE:ref1,IEEE:ref2}. Characteristics of chaos have been observed in many
experiments: externally driven and self-driven plasma systems. In driven systems chaotic
behavior in pulsed plasma discharge and period doubling cascade to chaos in thermionic
plasma discharge \cite{IEEE:ref3}, and in self-driven case i.e using no
external perturbation in the system, periodic to chaotic transition
have been observed\cite{IEEE:ref6,IEEE:ref7}. Experiments on plasma that exhibit period
doubling and chaos using external driver (oscillator) in a double plasma device has been done
by Ohno et al. \cite{IEEE:ref4}. The period subtracting phenomena,  has been observed in
externally driven ion-beam plasma system \cite{IEEE:ref5}. In this experiment
oscillation periods decreases in the sequence 6 period$\rightarrow$ 5 period $\rightarrow$ 4 period and
so on. Another phenomena in the driven plasma system is period adding opposite to the
period subtracting. Here periods successively increase with control parameters. %\cite{IEEE:ref8}.

In all the above mentioned experiments, the fluctuations are observed in the bulk region of the plasma,
where as in the present experiment we are reporting, the chaotic oscillation, period subtracting  and 
period doubling phenomena that were observed in the sheath region. The plasma sheath is a region with large electric fields where charged particles can encounter acceleration to high energies. This can give
rise to various types of fluctuations through wave particle interaction. These reasons motivated us
to investigate the fluctuations in the sheath region of a cylindrical dc glow discharge plasma. Moreover the sheath region is important from application point of view like material processing and dust levitation etc. Chaotic oscillations were observed during the initial phase of high pressure discharges. With increase in the discharge voltage it goes to an ordered state through period subtracting and finally the modes vanish and the plasma becomes stable via period doubling.

\section{Experimental Setup and Results}
\begin{figure}[htbp]

\includegraphics[width=8.5 cm]{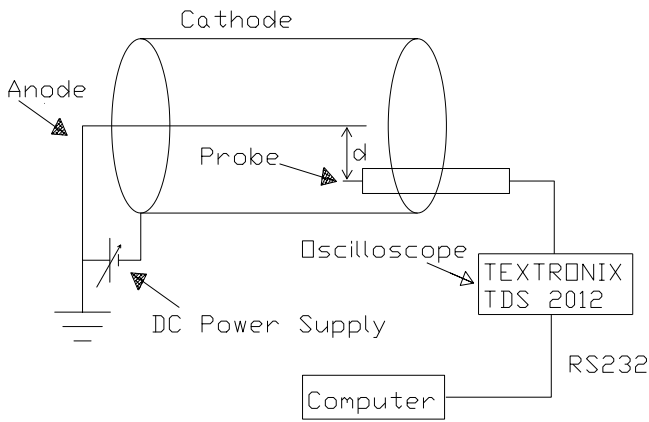}
\caption{Schematic diagram of Experimental setup of cylindrical dc discharge
plasma system with Langmuir probe fitted with Digital oscilloscope. A dc voltage has been applied
to the cathode (outer cylindrical surface) with respect to the grounded coaxial cylindrical rod and plasma is formed
inside it}
\label{fig:sys}
\end{figure}

The Experiment was carried out in a coaxial cylindrical dc glow discharge system with argon as shown in the figure~\ref{fig:sys}.
The hollow stainless steel (SS) outer cylinder of  45 mm diameter is the cathode and the SS rod  of 1 mm  diameter  inside the cathode is the anode, which is grounded. The Whole system has been placed in a vacuum chamber and
evacuated to  a base pressure of 0.001 mbar by means of a rotary pump. Argon gas is introduced using precision needle valve into the chamber. Neutral gas pressure has been maintained at a particular pressure and the discharge voltage to initiate a glow discharge plasma. A Langmuir probe made of tungsten has been used to measure floating potential fluctuations. It is of diameter 0.2 mm and length 4 mm, and movable along the plasma column. It is connected to the digital tektronix oscilloscope. The probe was placed in the sheath region i.e. about 0.5 cm from the cathode wall. Data has been transferred to the computer through the USB port.

\section{ Results and Discussions}
\label{sec:res}

\begin{figure}[htbp]
\includegraphics[width=8.5 cm]{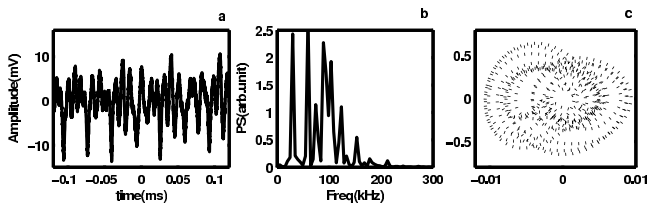}
\caption{$(a)$, $(b)$, $(c)$ represent the raw data, power spectrum(PS) and phase  space plot at 283 volts respectively. From power spectrum and phase space plot it is obvious that the signal is chaotic in nature.}
\label{fig:283}
\end{figure}

\begin{figure}[htbp]
\includegraphics[width=8.5 cm]{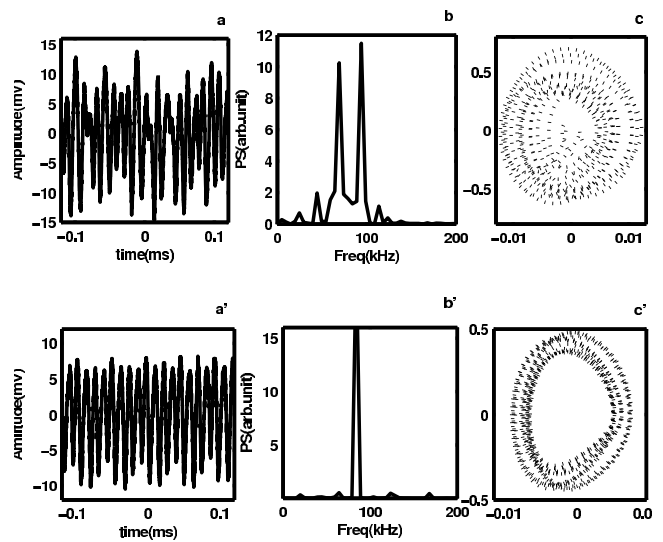}
\caption{(a), (b), (c) and $(a')$, $(b')$, $(c')$ are the raw data, power spectrum and phase space plot at 284 and 286 volts respectively. Through period subtraction the system attains five period at discharge voltage 286 volts from the seven period state at voltage 284 volts.}
\label{fig:284-286}
\end{figure}

\begin{figure}[htbp]
\includegraphics[width=8.5 cm]{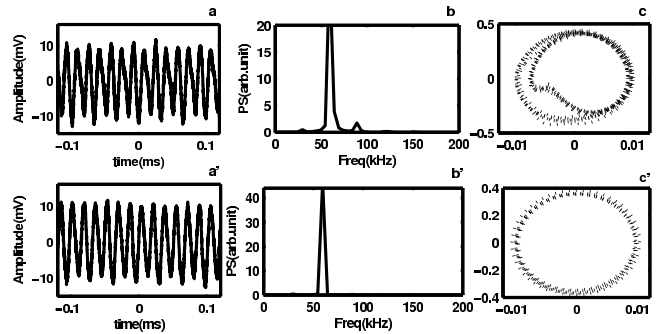}
\caption{In this figure in sequence $(a)\rightarrow(b)\rightarrow(c)$  the system goes to to state having three period at 288 volts from five period state at 286 volts (shown in ~\ref{fig:284-286}$(b')$ ) and in sequence $(a')\rightarrow(b')\rightarrow(c')$  to single period at 289 volts.}
\label{fig:288-289}
\end{figure}

Though plasma could be formed at low pressures, the chaotic oscillations in the floating potential were observed 
only above 0.4 mbar. The present experiments were  carried out at a neutral gas pressure of 0.95 mbar of Argon. At this particular pressure the discharge voltage between the anode and the cathode was increased slowly till the discharge was obtained  at  283 volts. At this voltage the fluctuating signal and its power spectrum are shown in the
figure~\ref{fig:283} (a) (raw data), (b) (corresponding power spectrum), and (c) phase space plot. The broad band power spectrum and phase space plot indicate the chaotic state at the initial stage of the dc discharge. On further increasing  the voltages we observed a period subtraction to take place and at 284 volts when two modes of frequencies 69.17, 93.87 kHz with maximum power, and 4.9, 24, 44.47, 113.6, and 123.5 kHz with smaller power shown in figure~\ref{fig:284-286}(b), i.e. seven periods appear. When the discharge voltage is 286 volts, by period subtraction the system reaches the state having five periods as shown in figure~\ref{fig:284-286}$(b')$. When the voltage was increased to 288 volts three modes of frequencies 29.64, 59.29 and 88.93 kHz, and at  289 volts only one period of frequency 59.29 kHz were seen as shown in the figure~\ref{fig:288-289}(b) and ~\ref{fig:288-289}$(b')$ respectively. These are also clear from the phase space plots of the data at the different voltages. From the respective phase space plots also, it is clear that the system goes from chaotic to coherent oscillations through period subtracting sequences.   

\begin{figure}[htbp]
\includegraphics[width=8.5 cm]{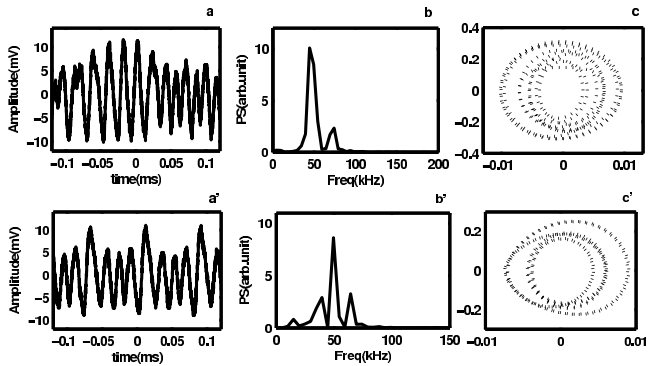}
\caption{ Further increase in voltage the system goes through two successive bifurcations.  The sequences $(a)\rightarrow(b)\rightarrow(c)$ and $(a')\rightarrow(b')\rightarrow(c')$ show mode appears at 290 and 291 volts respectively due to bifurcation. }
\label{fig:290-291}
\end{figure}

\begin{figure}[htbp]
\includegraphics[width=8.5 cm]{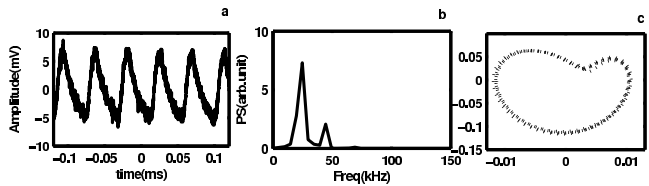}
\caption{$(a)\rightarrow(b)\rightarrow(c)$ are the 
raw data, Power spectrum and phase space plot at 292 volts. }
\label{fig:292}
\end{figure}

Further increase in discharge voltage  generates new modes and a period doubling is seen. It is clear from ~\ref{fig:290-291}$(b)$ that the power spectrum  at the 290 volts the frequencies are respectively 44.47 and 74.11 kHz.  Four new modes of frequencies 14.82 kHz, 39.53 kHz, 49.42 kHz, and 64.23 kHz emerge through period doubling at 291 volts as shown in the power spectrum ~\ref{fig:290-291}$(b')$. So again, the system evolved through period doubling in the following sequences- 1 period $\rightarrow$ 2 period $\rightarrow$ 4 period, for the discharge voltage sequence  289 $\rightarrow$ 290 $\rightarrow$ 291 volts. On further increasing the discharge voltages the system  generate new modes of frequencies 24.7 and 44.47 kHz through period subtraction at 292 volts as shown in the figure~\ref{fig:292}$(b)$. The new modes having almost halve of the frequencies that appears at 290 volts i.e. at the first bifurcation. On further increasing the voltages, all the instabilities vanishes, i.e. plasma floating potential becomes steady.
\begin{figure}[htbp]
\includegraphics[width=8.5 cm]{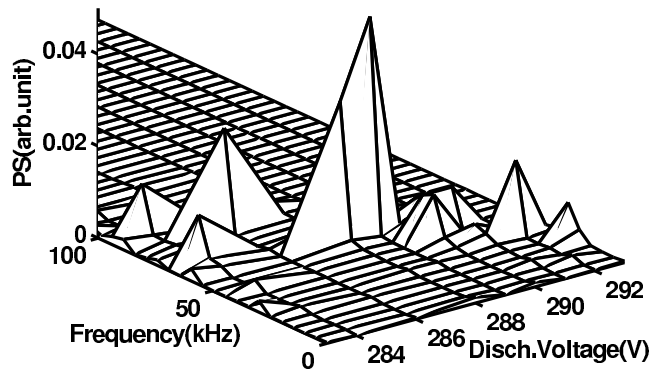}
\caption{Changes of  power and frequencies with discharge voltages.x-axis represents discharge potential,
y-axis frequencies, and z-axis represents power spectrum(PS) of the signals at different discharge voltage. With increase in the discharge voltage low modes appears.}
\label{fig:3d}
\end{figure}

The figure~\ref{fig:3d} shows the 3D plot of power spectrum of the data at different voltages. From this figure it is clear that with increase in the discharge voltage frequencies of the oscillation frequencies decrease.

An interesting phenomena is that an anode glow appears with the oscillations. Initially a spot 
like glow appears on the anode and with increasing voltage it spreads on the whole anode, and  with the  vanishing of the oscillations the glow also vanishes. So this can be related to anode glow instabilities and then transition
to stable potential structures like double layers etc.

Since the measurements were carried out in the cathode sheath regions, it is must likely that these 
oscillations could be due to ions that are accelerated within the cathode sheaths. The lower limit to
the frequencies observed would be of the order of inverse of ion transit time\cite{IEEE:ref10} 
across the sheath cathode sheath- 
%
%\begin{equation}
%\label{eqn:1}
%frac\tau=3{(\frac{eV_c}{kT_e})}^{1/4}\frac{1}{w_i} 
%\end{equation}   
% and the lower limit is given by the ion transit time 
\begin{equation}
\label{eqn:1}
\frac{1}{\tau}={\frac{1}{(\frac{eV_c}{kT_e})^{3/4}}}w_i
\end{equation}
where $V_c$, $T_e$ and $w_i$ are the the potential difference between plasma and cathode, plasma electron temperature and ion plasma frequency respectively.For the electron temperature $T_e\approx$ 2 eV, $V_c=$270 volts, this frequency is about 7.5 kHz. The upper limit would be of the order of ion plasma frequency, which is about 300 kHz for our parameters. The observed frequencies are from about 5- 170 kHz, which falls within the range that could be expected. 
              
As a part of our future plan, we intend to investigate the period subtraction and doubling phenomena in the bulk region of the plasma and also investigate their long range correlation.
%\newpage
\section*{Acknowledgment}
The authors would like to thank Ramesh Narayanan for useful discussion during the works.

\end{document}